\documentclass[aps,prl,twocolumn,showpacs,amsmath,amssymb,floatfix,
superscriptaddress]{revtex4}
\usepackage{graphicx}

\newcommand {\be}{\begin{equation}}
\newcommand {\ee}{\end{equation}}

\begin{document}

\title{Collective behavior of heterogeneous neural networks}
\date{\today}

\author{Stefano Luccioli}
\affiliation{Istituto dei Sistemi Complessi, Consiglio Nazionale
delle Ricerche, via Madonna del Piano 10, I-50019 Sesto Fiorentino, Italy}
\affiliation{INFN, Sez. Firenze, via Sansone, 1 - I-50019 Sesto Fiorentino, Italy}
\author{Antonio Politi}
\affiliation{Istituto dei Sistemi Complessi, Consiglio Nazionale
delle Ricerche, via Madonna del Piano 10, I-50019 Sesto Fiorentino, Italy}

\begin{abstract}
We investigate a network of integrate-and-fire neurons characterized by a
distribution of spiking frequencies. Upon increasing the coupling strength, the
model exhibits a transition from an asynchronous regime to a nontrivial
collective behavior. At variance with the Kuramoto model, (i) the macroscopic
dynamics is irregular even in the thermodynamic limit, and (ii) the microscopic
(single-neuron) evolution is linearly stable.
\end{abstract}


\maketitle

The investigation of networks of oscillators can provide new insights on the
basic mechanisms which underlie brain functioning. In particular, the
spontaneous onset of a collective dynamics is an intriguing phenomenon that can
contribute to information transmission across different brain areas. Given the
large number $N$ of neurons (oscillators) present in a real brain, it is
tempting to adopt a statistical-mechanics point of view and thereby investigate
the behavior for $N\to\infty$ (the so-called thermodynamic limit). Two different
setups are typically invoked \cite{GolombHanselMato_2001}: ({\it i}) sparse
networks, characterized by a fixed number of synaptic connections; ({\it ii})
massively connected networks, where the number of connections is proportional
to $N$. In the former case, the local field seen by the single neurons naturally
fluctuates even for $N\to\infty$ (being the sum of a fixed finite number of
different input signals), consistently with the experimental evidence of an
irregular background activity in the cerebral cortex~\cite{Destexhe_NatureNS2003}.
The latter setup has the advantage of being, at least in
principle, amenable to an exact mean-field treatment, although microscopic
fluctuations survive only if inhibition and excitation balance each
other~\cite{vanvreeswijk_sompolinsky_1998}.

In this Letter, we numerically show that an irregular microscopic and
macroscopic dynamics can generically arise even in an inhibitory, globally
coupled network. More precisely, we consider a heterogeneous network of
pulse-coupled integrate-and-fire (IF) neurons~\cite{tuckwell_1998}, each
characterized by a different bare spiking frequency. This setup is similar to
that of the Kuramoto model (KM) \cite{kura}, where each single oscillator is
identified by a phase variable $\phi$. The analogy is so tight that it has
even been shown that the pulse-coupling mechanism characterizing IF neurons
reduces, in the weak coupling limit, to that of the KM, the only difference
being that the coupling function is not purely sinusoidal~\cite{hansel_1995}. 
It is therefore quite important to clarify to what extent a network of IF
neurons reproduces the KM scenario for stronger coupling strengths, especially
by recalling that the KM is often invoked while testing new ideas on the
control of synchronization within neural contexts \cite{tass_95}. Finally, in
order to make the model closer to a realistic setup, we include delay to
account for the finite propagation time of the electric pulses.

Our strategy consists in studying the macroscopic collective dynamics in the
large $N$-limit, for different values of the coupling strength $g$. In the KM,
it is known that for a weak enough coupling, the single oscillators rotate
independently of each other. On the other hand, above a critical value, a
subset of them mutually synchronize, as signalled by a non zero value of the
order parameter $\rho = |\langle {\rm e}^{i\phi}\rangle|$ (the angular
brackets denote an average over the rotators).
In this Letter we show that IF neurons give rise to a similar but 
substantially different scenario. First of all, the (second) maximum
Lyapunov exponent is always negative~\cite{foot0}, implying that the evolution
must eventually converge to a periodic orbit. On the other hand, the study of
relatively small networks shows that the time needed to approach a periodic
orbit is exponentially long with the system size, implying that the
``transient" extends over increasingly longer time scales. In other words, this
is an instance of {\it stable chaos} \cite{poltorc}, a phenomenon already detected
in networks of pulse coupled oscillators without delay \cite{zillmer1},
although its onset in systems with delayed coupling is controversial
\cite{RudigerPRE2009,TimmeTOT_2008_2009}.

A second difference is that, at variance with the KM, the coupling contributes 
also to slowing down the spiking activity of the single neurons (a somehow
similar mechanism operates in ensemble of cold atoms \cite{JPP}) and drives a 
subset of neurons below the firing threshold -- a phenomenon reminiscent of 
oscillator-death \cite{Bressloff_Coombes2000}.
However, the most striking difference concerns the above-threshold regime,
as the overall neural activity is not {\it simply} periodically modulated,
but exhibits irregular, seemingly chaotic, oscillations (still in the presence
of a negative ``microscopic" second Lyapunov exponent). Nothing of this type
has been observed in the corresponding setup of a KM with delayed coupling
\cite{yeung_strogatz1999}.

The evolution equations for the $N$ membrane potentials $v_i$ write,
\be
\dot{v}_{i}=a_{i}-v_{i} - \frac{g}{N} \sum_{n|t_n<t} S_{i,l(n)}
\delta(t-t_{n}-t_d) 
\label{eq:model}
\ee
where all variables are expressed in adimensional units. When $v_{i}$ reaches
the threshold $v_{i}=1$, it is instantaneously reset to the value $v_{i}=0$,
while a spike is emitted (and received with a delay $t_d$). The network is
assumed to be heterogeneous, in that different neurons are exposed to
different suprathreshold currents $a_{i}$; $(\nu_0)_i =1/\ln[(a_{i}/(a_{i}-1)]$
is the bare spiking frequency. $S_{i,l}$ denotes the connectivity matrix and
the sum in Eq.~(\ref{eq:model}) runs over the spikes received by the neuron $i$. 
Finally, the coupling strength $g$ is our control parameter: the negative sign
in front of last term in the r.h.s. means that we assume inhibitory coupling.
Notice also that the same last term does not only couple the oscillators but
modifies also their frequency.

\begin{figure}[ht]
\begin{center}
\includegraphics[angle=0,width=7.5cm,clip]{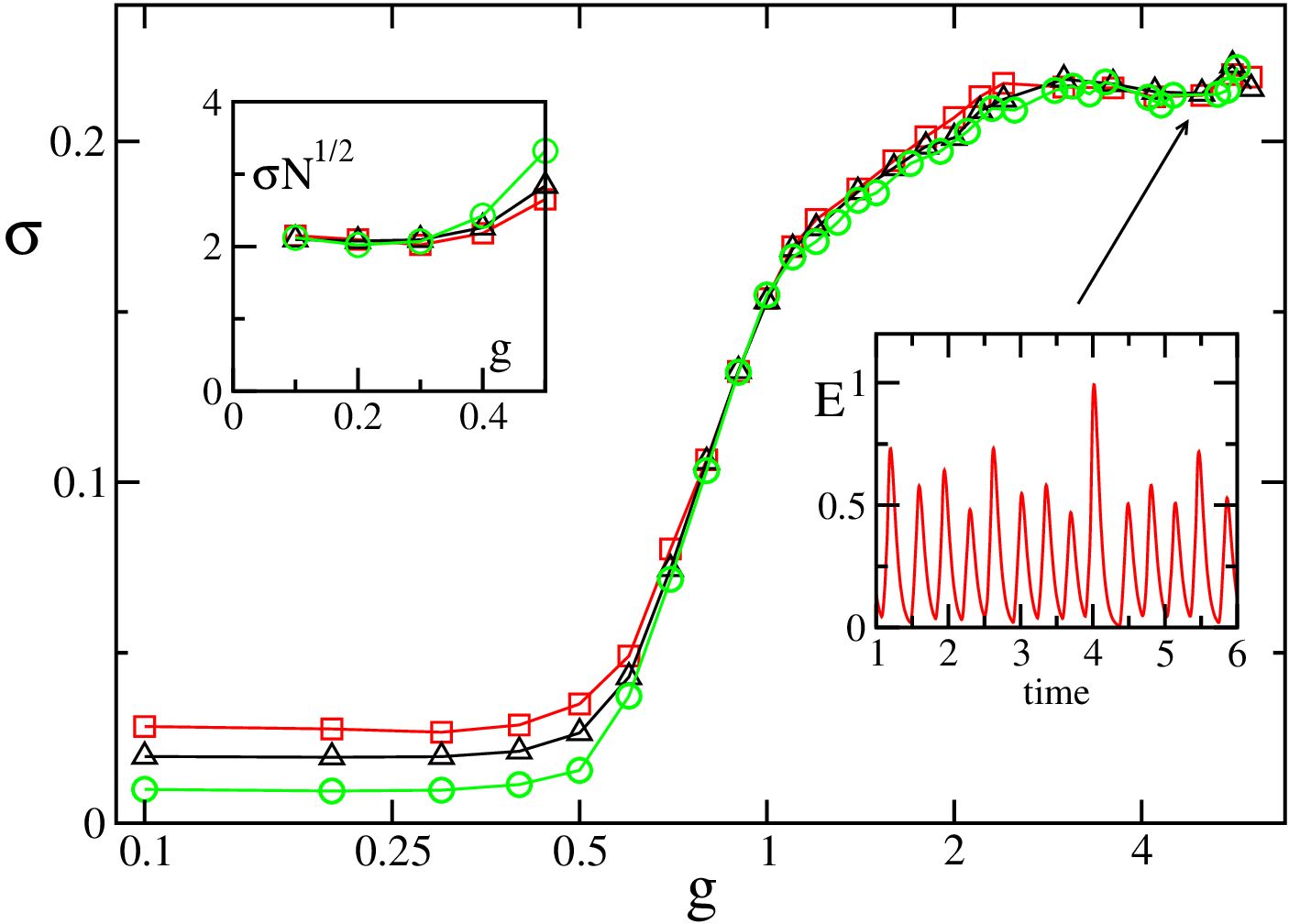}
\end{center}
\caption{(Color online) Standard deviation, $\sigma$, of the effective field $E$ 
as a function of the coupling strength $g$ for $N$=5,750, (red) squares, 
$N$=11,500, (black) triangles, and $N$=46,000, (green) circles. The upper inset
contains the rescaled standard deviation. The lower inset contains an instance
of the time evolution of $E$ for $g=5$ and $N=$5,750.}
\label{fig:diag}
\end{figure}

\begin{figure}[ht]
\begin{center}
\includegraphics[angle=0,width=7.5cm,clip]{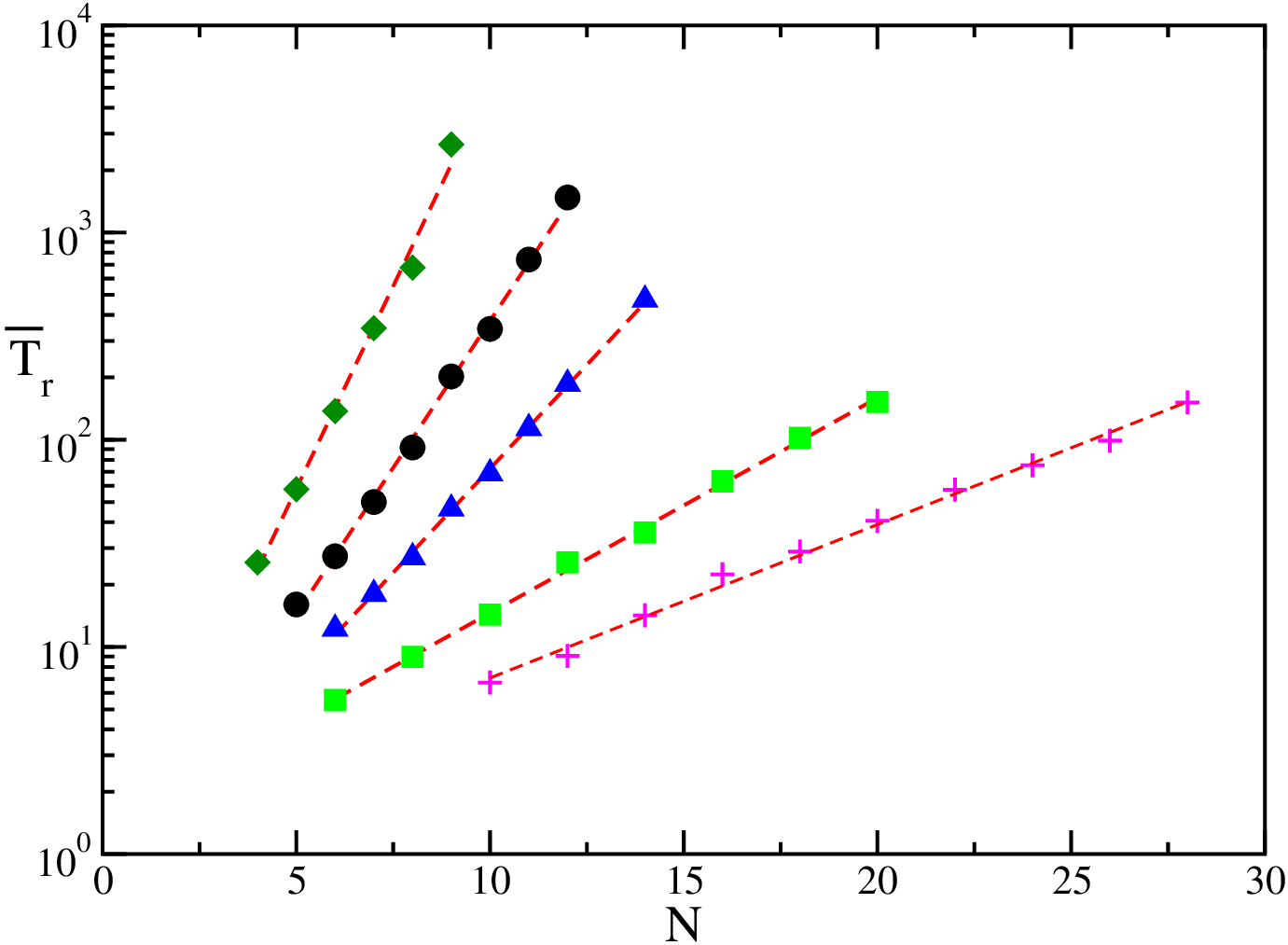}
\end{center}
\caption{(Color online) Average transient length ${\overline T}_{r}$ 
as a function of $N$ for $g=0.3$, 0.7, 1.3, 3, and 5 (diamonds,
circles, triangles, squares, and plusses, respectively).
The (red) dashed lines are exponential fits.}
\label{fig:trans}
\end{figure}

All the simulations reported in this Letter refer to a {\it globally coupled}
network, i.e. $S_{i,l}=1$ for any $i,l$, but we have verified that the
introduction of additional disorder (by randomly removing a fixed fraction of
connections) does not substantially modify the overall scenario. The delay is
set everywhere equal to $t_d = 0.1$, while the currents $a_{i}$ are randomly and
uniformly distributed in the interval $[1.2,2.8]$. These parameter values are
consistent with those selected in Ref.~\cite{RudigerPRE2009}, where they have
been chosen on the basis of biological motivations. 
In our case, the Kuramoto order parameter $\rho$ cannot be used to characterize
the onset of a collective dynamics. In fact, the inhibitory coupling may drive
the potential $v_i$ below the reset value, thus making the transformation of
the $v_i$ potential into a phase-like variable ill-defined. The difficulty can be 
overcome by coarse graining the spiking
activity. We do so
by {\it dressing} each spike with a finite width and thereby construct a
smooth effective field $E$. If we assume the pulse shape,
$p(t):= \alpha^2 t \exp(-\alpha t)$ ($t>0$), the corresponding
field $E$ can be generated by integrating the equation,
\be
\label{eq:effective}
\ddot E + 2\alpha \dot E + \alpha^2 E = \frac{\alpha^{2}}{N} 
\sum_{n|t_n<t} \delta(t-t_{n}-t_d)  \ . 
\ee
This procedure is often used to determine the field actually seen by
the single neurons~\cite{Abbott_VanVrewswijk1993}; here it is just a strategy to
construct a meaningful order parameter, that is defined as the standard
deviation $\sigma$ of $E$
($\sigma^2 = \langle E^2 \rangle_t - \langle E \rangle_t^2$, where 
$\langle \cdot \rangle_t$ denotes a time average). We choose $\alpha=20$, a
value that
corresponds to sufficiently broad pulses to get rid of the statistical
fluctuations, but not so large as to wash out the time evolution.
As long as the asymptotic regime is an asynchronous state characterized by a
constant activity, $\sigma$ is zero in the infinite $N$ limit, while any form of
collective dynamics gives rise to a nonzero $\sigma$. This is precisely what is
seen in Fig.~\ref{fig:diag}, where $\sigma$ is plotted versus the coupling
strength $g$ for different network sizes. Below $g_c\approx 0.5$, $\sigma$ is
quite small and appears to decrease as $1/\sqrt{N}$ with the system size (see
the left inset), indicating that the deviation from zero is a finite-size
effect. Above $g_c$, $\sigma$ starts to grow and is independent of the system
size, suggesting the onset of some form of synchronization (the right inset
contains an instance of the field evolution for $g=5$). Superficially, this
scenario is reminiscent of the synchronization transition observed in the KM.
In the following we show that there are several
conceptually relevant differences.
The first difference concerns the microscopic (single neuron) behavior. The
maximum Lyapunov exponent, $\lambda$, of the Poincar\'e map (to get rid of the
first zero Lyapunov exponent) is negative both below and above the transition
and does not depend on $N$ for large system sizes~\cite{foot1}. Altogether, the
stable microscopic dynamics observed in this setup
contrasts with the weakly unstable dynamics observed in the KM, where the
maximum Lyapunov exponent is positive, though scales as $1/N$ \cite{maistrenko}.
On the other hand, the transient time $T_r$ needed for a generic trajectory to
converge to some periodic orbit grows exponentially with $N$. This is
illustrated in Fig.~\ref{fig:trans}, where the average ${\overline T}_r$
(over more than 100 realizations of the disorder) is plotted
for different coupling strengths. There, one can also appreciate that
the exponential growth rate decreases systematically with increasing $g$.
Therefore, for large $N$, the relevant dynamical regime is represented by
the ``transient" dynamics, rather than by the periodic orbit approached over
astronomical time scales. This {\it stable chaos} scenario was first observed 
in the absence of delay~\cite{zillmer1} for networks of identical oscillators,
when disorder in the connectivity matrix is included. Its occurrence in the
presence of delay is somehow controversial. It appears that the length of the
transient depends crucially on the balance between the amplitude of the
effective disorder and the stability of clustered states. Whenever local
fluctations decrease with $N$, the transient length does not only stops
growing exponentially, but even decreases, since generic trajectories rapidly
approach one of the clustered states \cite{foot2}. At variance with the
previously considered setups~\cite{TimmeTOT_2008_2009,RudigerPRE2009}, the
disorder induced by the heterogeneity of the currents, survives in the
thermodynamic limit. Accordingly, the exponential growth of the transient
length is expected to persist for arbitrarily large $N$ even in the absence of
disorder in the connections. In fact, we find no evidence of a convergence
towards more coherent states.

\begin{figure}[ht]
\begin{center}
\includegraphics*[width=7.5cm,angle=0,clip]{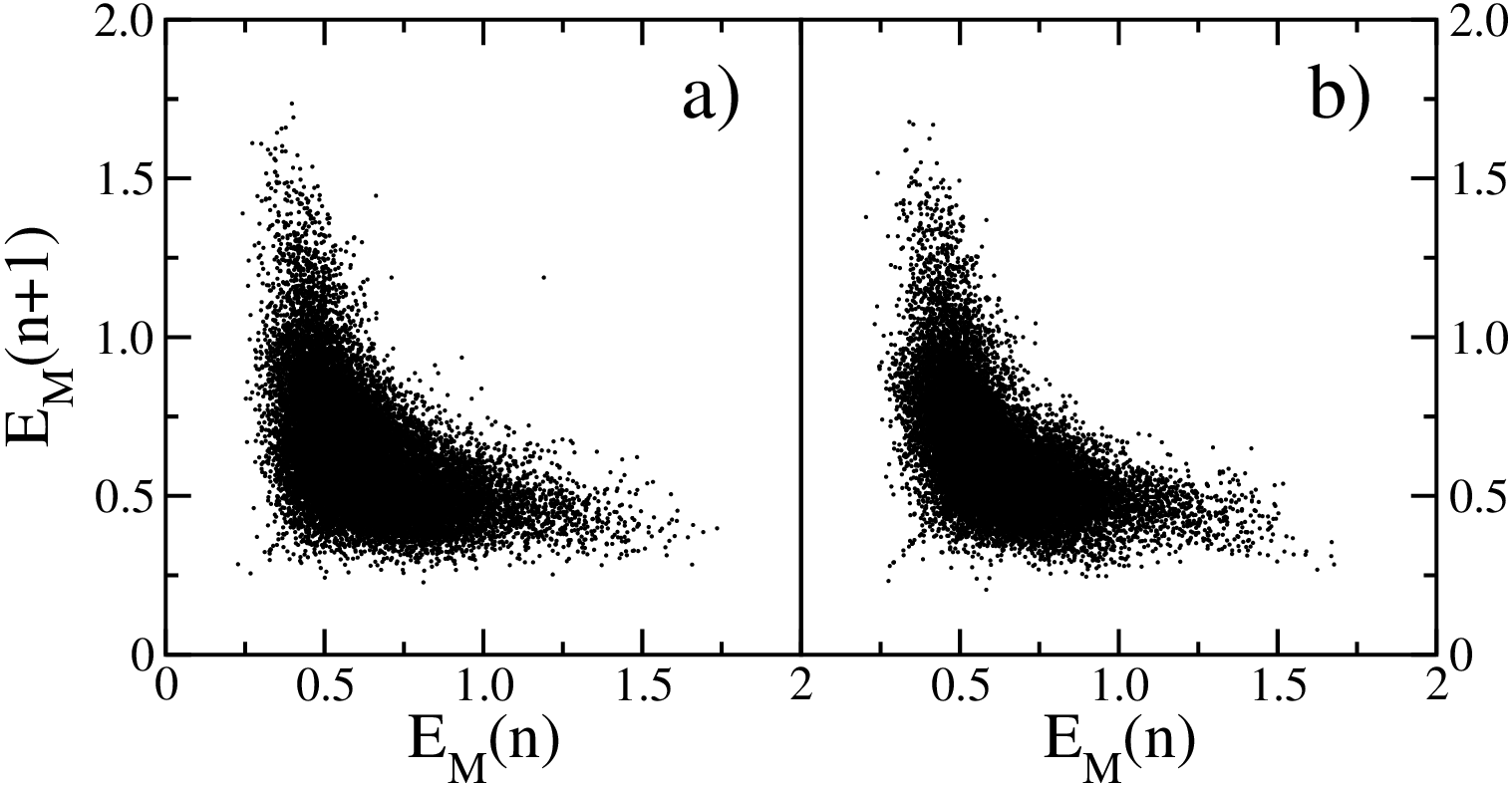}
\end{center}
\caption{Return maps for the maxima of the effective field $E$ 
when $g=5$ for $N$=11,500 (a) and $N$=46,000 (b).}
\label{fig:poin}
\end{figure}

The standard deviation $\sigma$ allows identifying the very existence of
collective fluctuations, but does not tell us anything about their dynamical
character. Up to $g \approx 2$, simulations performed for increasing $N$ suggest
that the field $E$ behaves periodically in the thermodynamic limit. On the
other hand, the right inset in Fig.~\ref{fig:diag}, which refers to $g=5$, reveals
a rather irregular behavior still for $N=46,000$. A more accurate analysis is
however necessary before making any claim. As a first test, we construct
a return map by plotting the $(n+1)$-st maximum $E_M(n+1)$ of the field versus
the previous one. In Fig.~\ref{fig:poin}, we see that the points in the
Poincar\'e section fill a broad and almost the same area for both $N=$11,500
and 46,000. Such features consistently indicate that the collective dynamics is
characterized by complex oscillations.

\begin{figure}[ht]
\begin{center}
\includegraphics[angle=0,width=7.5cm,clip]{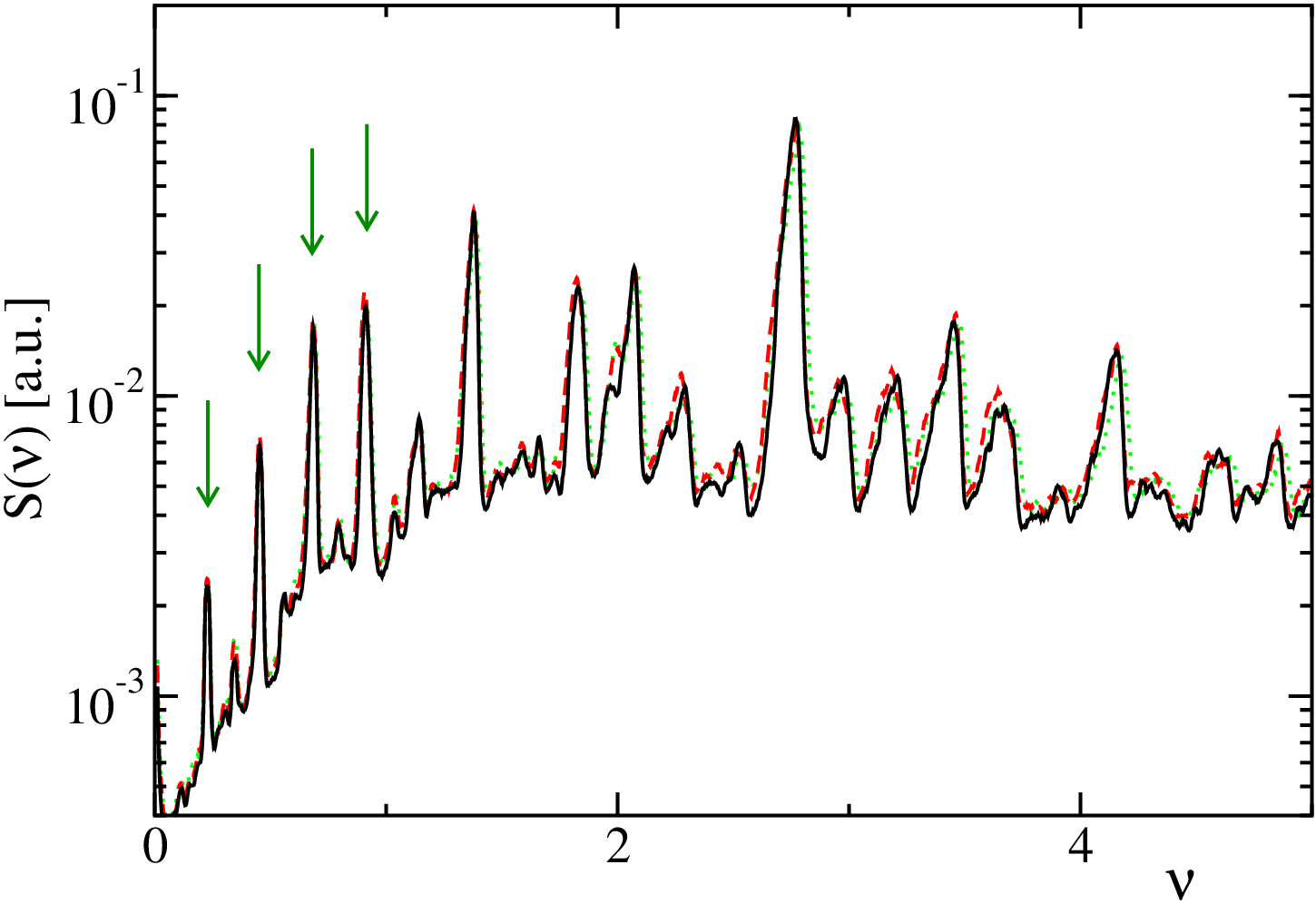}
\end{center}
\caption{(Color online) Power spectrum $S(\nu)$ of the effective field for
$N=5,750$, (dotted line) 11,500 (dashed line) and 46,000 (solid line). The
spectra have been obtained by Fourier transforming a signal of temporal
length $\approx 49,300$.  The arrows point to the frequencies identified in 
Fig.~\ref{fig:freq}}
\label{fig:spett}
\end{figure}

Next we characterize the collective motion by computing
the Fourier power spectrum $S(\nu)$ of the field $E$. The spectra reported in
Fig.~\ref{fig:spett} reveal several broad peaks whose width does not appear
to decrease for increasing $N$. This confirms
that the irregularity of the collective dynamics persists in networks of
arbitrary size and therefore differs from the periodic oscillations reported,
e.g. in Refs.~\cite{VanVreeswijk_1996,brunel_hakim1999}.

\begin{figure}[ht]
\begin{center}
\includegraphics[angle=0,width=7.cm,clip]{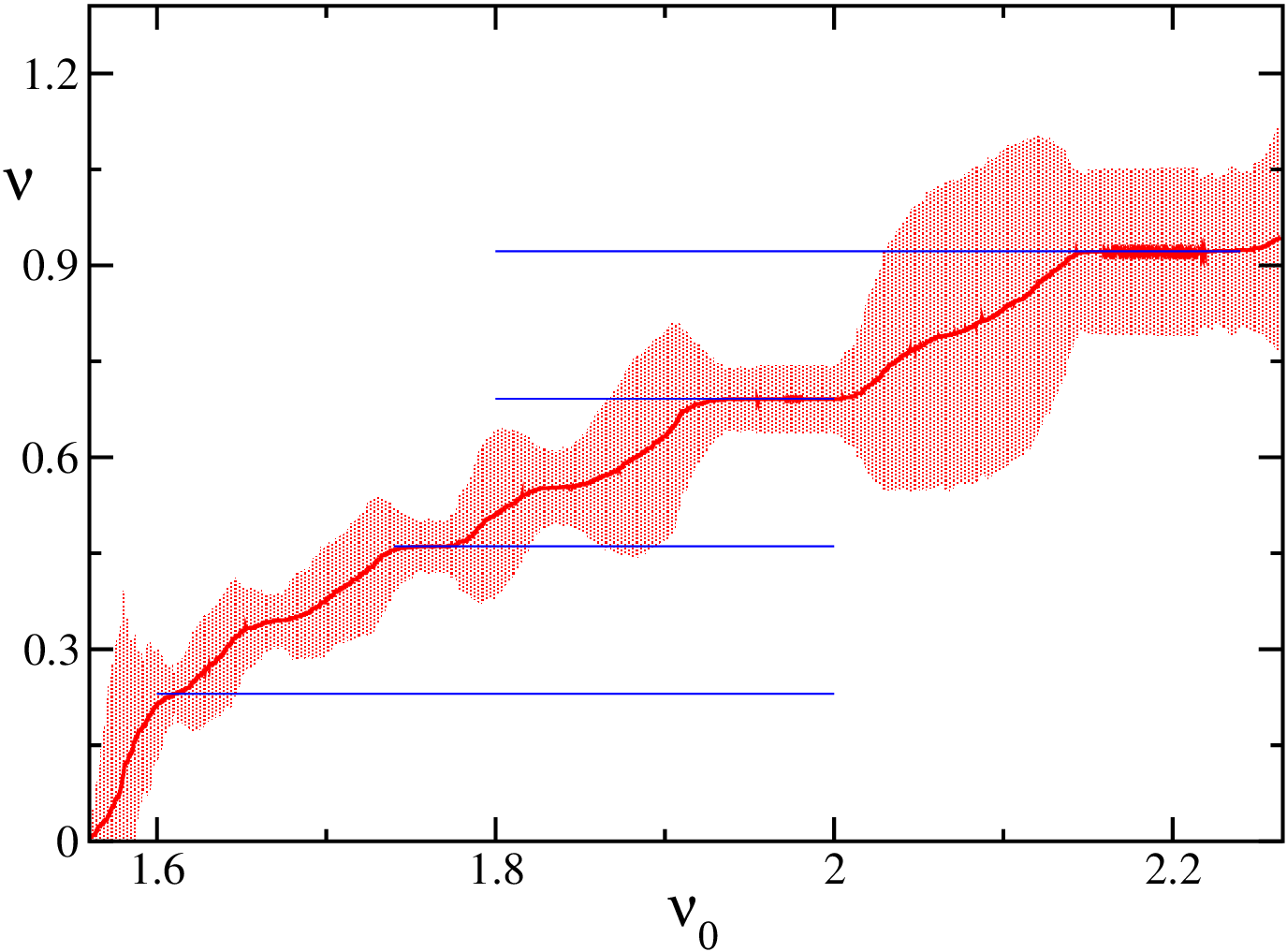}
\end{center}
\caption{(Color online) Average spiking frequency of the oscillators as a
function of the bare frequency (ranging in the interval $[0.558,2.26]$),
for $g=5$ and $N=$46,000 neurons (red solid line). The thin horizontal
lines are located at multiples of $\nu=0.23$. The shaded area identifies the
region covered by frequency fluctuations (see the text for further details).}
\label{fig:freq}
\end{figure}

In order to shed further light on the system evolution, we have analysed the
single-neuron behavior too. In Fig.~\ref{fig:freq}, the spiking frequency $\nu$
(defined as the inverse of the average inter-spike interval -- ISI) of the
single neurons is plotted versus the bare frequency $\nu_0\in[0.558,2.26]$
(again for $g=5$). The effective frequency is systematically smaller than
$\nu_0$; this is because the inhibitory coupling lessens the neural activity.
In fact, inhibition is so strong, as to bring the least active neurons below
threshold: neurons with $\nu_0<\approx 1.56$ do not fire at all, and thus do not
actively contribute to the network dynamics: they are just slaved by the other
degrees of freedom. 

The appearance of plateaux (the widest ones corresponding to harmonics
of the frequency $\nu=0.23$) reveals that neurons with similar bare frequencies
lock together, as it is naturally expected for periodically forced oscillators
(see the phenomenon of Arnold tongues). However, in this case, the forcing
field is not periodic: the shaded region around the curve $\nu(\nu_0)$
highlights the fluctuations of the ISI (its vertical width is equal to three
standard deviations of $\nu$). We have verified that such a width
does not vanish upon increasing $N$, while the neurons within the same plateau
are frequency- but not phase-locked.
It is tempting to trace back the irregular collective motion to the presence
of neurons that are nearly at threshold, whose activity is quite sporadic.
However, we have verified that the overall evolution is almost unchanged when
such neurons (and those which do not spike at all) are removed from the
outset.

All of our numerical simulations confirm that the irregularity of the
collective dynamics persists for $N \to\infty$. It is important to
realize that this scenario is a priori legitimate, since the dynamics is ruled,
in the thermodynamic limit, by a suitable nonlinear functional equation. In
this case, the relevant object is the probability density $P(v,\nu_0,t)$ for the
membrane
potential of the neurons, whose bare spiking frequency lies in the interval
$[\nu_0,\nu_0+d\nu_0]$ to belong to the interval $[v,v+dv]$ at time $t$. As
functional equations involve infinitely many degrees of freedom, one can,
in principle, expect an arbitrary degree of dynamical complexity.
In models such as the networks considered in~\cite{vanvreeswijk_sompolinsky_1998},
the corresponding probability density is a Gaussian and it is therefore
described by just two variables. As a result, in that context one
cannot observe anything more complex than periodic oscillations. In the
standard KM it has been proved that not even periodic oscillations can arise;
a periodic collective motion can be observed only by invoking a more complicate
nonlinear dependence on the order parameter~\cite{pikovsky}. On the other hand,
globally coupled logistic maps exhibit an infinite dimensional dynamics
\cite{kaneko}. The problem of determining the active modes in a globally coupled
system is, in general, hard to solve, as the modes may be highly singular and it
might not be obvious which basis to use to expand the functional equation. In
the context of the model studied in this Letter, we face the additional
difficulty that the microscopic dynamics is characterized by a negative
Lyapunov exponent and there is no guarantee that the distribution
$P(v,\nu_0,t)$ is smooth.

We wish to thank A.~Torcini, Th.~Kreuz and S.~Olmi for enlightening discussions.
We acknowledge financial support from the CNR project ``Dinamiche cooperative in
strutture quasi uni-dimensionali".


\begin{thebibliography}{xx}
\bibitem{foot0} As long as the asymptotic state is not a fixed
point, the maximum exponent cannot be strictly negative.
\bibitem{foot1} After averaging over different realizations of the
disorder, we find $\lambda =-0.105 \pm 0.002$ for $g=0.2$, 
$\lambda= -0.253 \pm 0.002$ for $g=0.7$ and $\lambda= -0.572 \pm 0.003$ for $g=5$.
\end{thebibliography}

\end{document}